\documentclass[fleqn,usenatbib]{mnras}

\usepackage{newtxtext,newtxmath}

\usepackage[T1]{fontenc}

\DeclareRobustCommand{\VAN}[3]{#2}
\let\VANthebibliography\thebibliography
\def\thebibliography{\DeclareRobustCommand{\VAN}[3]{##3}\VANthebibliography}

\usepackage{graphicx}	
\usepackage{amsmath}	
\usepackage[normalem]{ulem}
\usepackage{subfig}

\title[X-ray polarization of LMC X-1]{The first X-ray polarimetric observation of the black hole binary LMC X-1}

\author[{\em IXPE} Collaboration]{
J. Podgorn\'y,$^{1,2,3}$\thanks{E-mail: jakub.podgorny@astro.unistra.fr} L. Marra,$^{4}$ F. Muleri,$^{5}$ N. Rodriguez Cavero,$^{6}$ A. Ratheesh,$^{5}$  M. Dov{\v{c}}iak,$^{2}$  \newauthor \; R. Miku{\v{s}}incov\'a,$^{4}$ M. Brigitte,$^{2}$ J. F. Steiner,$^{7}$ A. Veledina,$^{8,9}$
% ...
\\
\\
     Affiliations are listed at the end of the paper. 
}

\date{Accepted XXX. Received YYY; in original form ZZZ}

\pubyear{2023}

\begin{document}
\label{firstpage}
\pagerange{\pageref{firstpage}--\pageref{lastpage}}
\maketitle

\begin{abstract}
We report on an X-ray polarimetric observation of the high-mass X-ray binary LMC X-1 in the high/soft state, obtained by the \textit{Imaging X-ray Polarimetry Explorer} ({\em IXPE}) in October 2022. The measured
polarization is below the minimum detectable polarization of 1.1 per cent (at the 99 per cent confidence level). Simultaneously, the source was observed with the {NICER}, {\em NuSTAR} and {\em SRG}/{ART-XC} instruments, which enabled spectral decomposition into a dominant thermal component and a
Comptonized one. The low 2--8 keV polarization of the source did not allow for strong constraints on
the black-hole spin and inclination of the accretion disc. However, if the orbital inclination of about
36 degrees is assumed, then the upper limit is consistent with predictions for pure thermal emission
from geometrically thin and optically thick discs. Assuming the polarization degree of the Comptonization component to be 0, 4, or 10 per cent, and oriented perpendicular to the polarization of the disc emission (in turn assumed to be perpendicular to the large scale ionization cone orientation detected in the optical band), an upper limit to the polarization of the disc emission of 1.0, 0.9 or 0.9 per cent, respectively, is found (at the 99 per cent confidence level).
\end{abstract}




\section{Introduction}\label{introduction}

LMC X-1 is the first discovered extragalactic black-hole (BH) X-ray binary system \citep{Mark1969}. Being located in the Large Magellanic Cloud, the source has a well determined distance of $50 \pm 1$ kpc \citep{Pietrzynski2013}. LMC X-1 is persistent and bright; hence, it has been studied extensively since its discovery. X-ray binary systems typically exhibit two distinct spectral states in the X-ray band: the `high/soft state' in which the thermal emission from a multi-temperature blackbody accretion disc \citep{Shakura1973, Novikov1973} is dominant and the `low/hard state' in which a power-law component is dominant \citep{Zdziarski2004, Remillar2006}. While many X-ray binary systems change their spectral state over time, LMC X-1 has always been observed in the soft state with $L_\textrm{X} \sim 2 \times 10^{38}$ erg\,s$^{-1}$ \citep{Nowak2001, Wilms2001}. Typically more than 80 per cent of the X-ray flux can be attributed to the thermal/disc component \citep[see e.g.][]{Nowak2001, Steiner2012, Bhuvana2021, Jana2021, Bhuvana2022}. The remainder of the X-ray flux can be decomposed into coronal power-law emission \citep{Sunyaev1980}, a broad Fe-line from the relativistic disc \citep{Fabian1989}, and a narrow Fe-line that most likely originates from scattering off highly ionized wind from the stellar companion \citep{Steiner2012}.

Optical and near-infrared observations reveal an O7/O9 giant donor with a mass of $M_2 = 31.8 \pm 3.5$\,M$_{\odot}$ \citep{Orosz2009}. The same dynamical study confirms a BH accretor with a mass of $M_{\textrm{BH}} = 10.9 \pm 1.4$\,M$_{\odot}$ and an orbital inclination $i = 36.4\degr \pm 1.9\degr$. The measured orbital period of LMC X-1 is $3.90917\pm  0.00005$~days \citep{Orosz2009}, based on high-resolution optical spectroscopy. Over an orbit, the X-ray flux exhibits achromatic sinusoidal amplitude modulations of 7 per cent associated with the inferior/superior conjunctions and Thomson scattering and absorption by the stellar wind \citep{Nowak2001, Orosz2009, Hanke2010}. 
Strong red noise variability is observed on timescales shorter than the orbital period \citep{1999AJ....117.1292S, Nowak2001, Bhuvana2022}. Also, low frequency quasi-periodic oscillations (QPOs) were observed on several occasions \citep{1989PASJ...41..519E, Alam2014}, which do not fit well within the standard low-frequency QPO ABC classification \citep{Casella2005}.

Measurement of the BH spin in LMC X-1 is of great interest. The system is a high-mass X-ray binary, and estimation of the BH spin is useful for stellar evolution and cosmological studies \citep[see e.g.][]{Qin2019, Mehta2021}. The donor star is 5~Myr past the zero-age main sequence and believed to be filling 90 per cent of its Roche lobe. This, and the inferred dynamical parameters of the system, suggest that LMC X-1 is likely a precursor of an unstable mass transfer phase and a common-envelope merger \citep{Podsiadlowski2003, Orosz2009, Belczynski2012}. Such systems are of potential interest for gravitational-wave studies, especially regarding the spin of the BH \citep{Belczynski2021, Fishbach2022, Shao2022}. Many spectroscopic studies have estimated the spin of the BH in LMC X-1, using the continuum and relativistic line fitting techniques in Kerr spacetime, assuming the spin is aligned with the system axis of symmetry \citep[see][for LMC X-1 studies beyond the Kerr metric]{Tripathi2020}. They infer remarkably high spin values: $0.85 \lesssim a \lesssim 0.95$ \citep[continuum method;][]{Gou2009, Mudambi2020, Jana2021, Bhuvana2022} and $0.93 \lesssim a \lesssim 0.97$ \citep[Fe-line method;][]{Steiner2012, Bhuvana2022}. Along with the high spin, high accretion rates of $0.07 \lesssim \dot{M}/\dot{M}_{\textrm{Edd}} \lesssim 0.24$ and luminosities $0.1 \lesssim L_{\textrm{X}}/L_{\textrm{Edd}} \lesssim 0.16$ are estimated \citep[the quantities are defined in][]{Bhuvana2022}. The power-law index tends to be steep $2 \lesssim \Gamma \lesssim 4$ \citep{Gou2009, Jana2021, Bhuvana2022, Nowak2001}. A counter-argument to the high spin of LMC X-1 through X-ray spectroscopy was given by \cite{Koyama2015} that introduced a double Compton component model to fit the data, which allows a larger disc inner radius, leading to a lower spin estimate.

X-ray polarimetry can constrain the geometry of the unresolved inner accretion flow and the inclination of the accretion disc with respect to the observer. It can also independently constrain the spin of the BH \citep{Connors1977, Stark1977, Connors1980, Dovciak2004, Dovciak2008, Li2009, Schnittman2009, Schnittman2010, Cheng2016, Taverna2020, Taverna2021, Krawczynski2022b}, especially in the high/soft state when the accretion disc is widely believed to extend to the inner-most stable circular orbit.

We present the first X-ray polarimetric measurement of LMC X-1, which serves as an example of an accreting BH caught in the thermal state. The \textit{Imaging X-ray Polarimetry Explorer} ({\em IXPE}) \citep{ixpe} observed LMC X-1 in the 2--8 keV band in which the disc emission dominates during October 2022. Simultaneous X-ray observations were performed with the {NICER} \citep{nicer}, {\em NuSTAR} \citep{nustar} and {ART-XC} \citep{2021A&A...650A..42P} instruments to better characterize the source spectrum. The {\em IXPE} observation of LMC X-1 helps fill out the sample of accreting BHs with X-ray polarization measurements which includes the accreting stellar-mass BHs in Cyg X-1 \citep{Krawczynski2022} and Cyg X-3 \citep{Veledina2023} (in the low/hard or intermediate states), and the supermassive BHs in MCG 05-23-16 \citep{Marinucci2022} and the Circinus galaxy \citep{Ursini2023}. We obtained a low upper limit on the 2--8 keV polarization of LMC X-1 in the thermal state. Our careful spectro-polarimetric analysis leads to constraints on polarization of the distinct X-ray spectral components and validates long-standing theoretical predictions for X-ray properties of the inner-most regions of accreting BHs.

Independent constraints on the accretion disc orientation are important when interpreting the X-ray polarization results. A $\sim 15$~pc parabolic structure in the form of a surrounding nebula (wind or jet powered) was detected in both optical and radio observations \citep{Pakull1986, Cooke2008, Hyde2017}. The nebula is aligned with an inner $\sim 3.3$ pc ionization cone of 50\degr projected full opening angle seen in He II and [O III] lines, which is believed to be directly related to the BH accreting structure \citep{Cooke2007, Cooke2008}. We use this large-scale measurement of the disc orientation to assess the X-ray polarization position angle measured by {\em IXPE} at sub-pc scales; this is similar to comparison made for Cyg X-1 \citep{Krawczynski2022}. The jet of LMC X-1 has not been detected yet \citep{Fender2006, Hughes2007, Hyde2017} and is likely to be switched-off since the binary is persistently in the thermal state \citep{Cooke2007}.

The paper is organized as follows. Section~\ref{observations} describes the observations and the data reduction techniques. Our spectral and polarimetric results are presented in Section \ref{analysis}. Section \ref{conclusions} provides a summary.

\section{Observations and data reduction}\label{observations}

{\em IXPE} \citep{ixpe} observed LMC~X-1 between 2022 Oct 19 15:01:48 UTC and 2022 Oct 28 04:39:09 UTC, under the observation ID~02001901 and for a total livetime of $\sim$ 562~ks for each of its three telescopes. Processed, Level~2, data already suitable for scientific data analysis were downloaded from the {\em IXPE} HEASARC archive.\footnote{Available at \url{https://heasarc.gsfc.nasa.gov/docs/ixpe/archive/}.}
Source and background regions were spatially selected in the {\em IXPE} field of view defining different concentric regions, both centered on the image barycenter. The source region is defined as a circle with radius 1.5 arcmin, while the background region is an annulus with inner and outer radii of 2.5 and 4 arcmin, respectively. We show these regions on top of the {\em IXPE} count maps in Appendix \ref{observations_remaining}.

Two different approaches were used to estimate the X-ray polarization. The first relies on the use of forward-folding fitting software \citep[we used \textsc{xspec},][version 12.13.0]{Arnaud1996} to model Stokes spectra $I$, $Q$ and $U$. This allows us to model the spectrum of the source $I$ with different components, associating to each of them a polarization model which is constrained using the $Q$ and $U$ spectra. An alternative approach makes use of the \textsc{ixpeobssim} package \citep{Baldini2022}, which provides tools for {\em IXPE} data analysis including the {\tt PCUBE} algorithm of the {\tt xpbin} function, which calculates the polarization degree and angle from the Stokes parameters without making any assumption on the emission spectrum. 
For \textsc{xspec} analysis, we used the formalism from \citet{stro17} and used the weighted analysis method presented in \citet{DiMarco2022} (parameter {stokes=Neff} in {\tt xselect}).

The polarization cubes (PCUBEs) for both the source and background regions generated with \textsc{ixpeobssim} combine the observations from each detector unit (DU), and return the total polarization degree and angle as well as the minimum detectable polarization (MDP) at 99 per cent confidence level. Using \texttt{xpbin} with the \texttt{PHA1, PHA1Q, PHA1U} algorithms, we created spectral files of Stokes $I$, $Q$ and $U$ parameters, respectively. These files are produced in the OGIP, type 1 PHA format, which is convenient for spectral, polarimetric, and joint analysis within \textsc{xspec}.

Appendix \ref{observations_remaining} contains a full description of the {NICER}, {\em NuSTAR} and {ART-XC} observations and the data reduction. This includes discussion of our use of the cross-calibration model {\tt MBPO} employed to reconcile discrepancies between the instruments and of level of the systematic uncertainties of the instruments.

\section{Data analysis}\label{analysis}

\subsection{Spectral and timing analysis}\label{spectral_analysis}

\begin{figure}
    \centering
    \includegraphics[width=8.8cm]{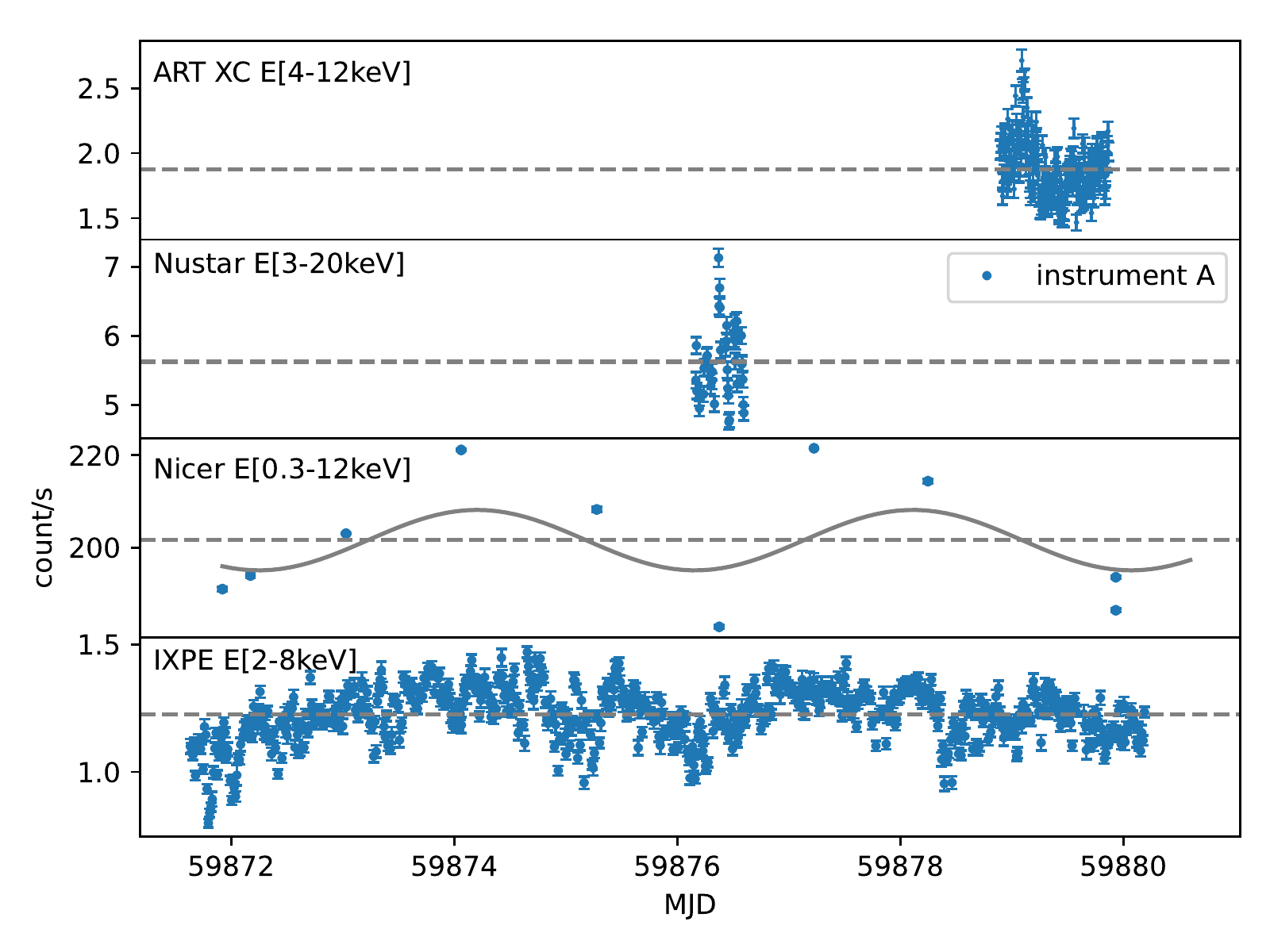}
    \caption{X-ray light curves of LMC X-1. \textit{Top panel}: {ART-XC} light curve for the energy range 4--12 keV. \textit{Second panel}: {\em NuSTAR} light curve for the energy range 3--20 keV from the instrument A of {\em NuSTAR}. \textit{Third panel}: {NICER} light curve for the energy range 0.3--12 keV with a sinusoidal curve showing the expected orbital modulations of the X-ray flux based on previous {\em RXTE} monitoring of the source. \textit{Bottom panel}: {\em IXPE} light curve for the energy range 2--8 keV. The dashed horizontal lines are the average count rate for each light curve.}
    \label{fig:lightcurves}
\end{figure}

Daily monitoring by the Gas Slit Camera (GSC) onboard of {MAXI} \citep{Maxi_2009} confirmed that during our observations, there were no outbursts or long-term flux variations that would suggest that the source departed from the high/soft state.
In this study, we analyzed in more detail the flux variability of LMC X-1 during the {\em IXPE} observation, using light curves from the simultaneous observations by {NICER}, {\em NuSTAR} and {ART-XC} (see Fig. \ref{fig:lightcurves}). We used the following energy ranges for the light curves: 0.3--12 keV, 3--20 keV and 4--12 keV, respectively for {NICER}, {\em NuSTAR} and {ART-XC}. Despite {ART-XC} registers useful signal up to 35~keV, we used shorter energy band for the timing analysis due to the sharp decrease of the mirror systems effective area above the nickel edge at $\approx 12$~keV. The corresponding time bins were 920 s for {NICER}, 400 s for {\em NuSTAR} and {ART-XC}, and 1000 s for {\em IXPE}.

The {\em IXPE} and {NICER} observations cover a period of 10 days, while {\em NuSTAR} and {ART-XC} complement these observations with  snapshots in the hard X-ray band. Our {\em IXPE} and {NICER} observations thus include about two and half orbits of the BH and companion star. \citet{Orosz2009} measured orbital modulations of the X-ray flux to be consistent with the periodicity measured from optical data. The X-ray orbital modulation was revealed via a set of {\em RXTE}/ASM \citep{rxte} data from over 12 years monitoring, and it was attributed to the electron scattering and absorption in the stellar wind from the companion star \citep{Orosz2009, Levine2011}.
To estimate the X-ray flux orbital modulations in the current observations, we took the orbital ephemeris from the `adopted' model in table~3 of \citet{Orosz2009}; in particular, we assumed the orbital period of 3.90917~days and the time of the superior conjunction of 53390.8436~MJD (Modified Julian Date).
We took the parameters of the best-fitting sinusoidal curve from their table~1 for the 1.5--12 keV energy band, where they reported parameters averaged over the 12 years observation with {\em RXTE}, and we rescaled to the {NICER} count rate. The {NICER} count rate versus orbital phase is then $f(\phi) = a_0 - a_1 \cos(2\pi\phi)$ where, once rescaled, the parameters are $a_0 = 201.69$ and 
$a_1=a_{1, \rm{RXTE}} \times \frac{a_0}{a_{0, \rm{RXTE}}}=6.51$ and $\phi$ is the phase.
The curve is shown along with the {NICER} data in the third panel of Fig.~\ref{fig:lightcurves}.

Comparison of the curve and the data indicates that the X-ray variations in the {NICER} light curve can be well explained by the expected orbital modulations. The amplitude of the {NICER} data modulation is higher than the amplitude from the {\em RXTE}-ASM analysis. This is most likely due to a different sensitivity of the instruments, the {NICER} camera being more sensitive in the low energies where most counts are detected and affected by the circumstellar absorption, and thus the amplitude of the modulation might be larger. Any stochastic variations, which can also contribute to the single observation, are smeared out in the averaging over 12 years of monitoring with {\em RXTE}. Similar modulations are apparent in the {\em IXPE} light curve. The X-ray flux minima correspond to superior conjunctions of the BH that are associated with the enhanced absorption and reduced scattered emission due to the wind from the companion.

In the light curves acquired in the hard X-ray band ({ART-XC} and {\em NuSTAR}), stochastic noise dominates over the orbital modulations. Similar to previous research \citep[see][]{Koyama2015}, we observe an increase in stochastic red noise variability with energy.  The power spectrum in the hard band can be described with a power law with index $\approx -1$ and normalization consistent with the previous measurements \citep[see e.g.][]{Bhuvana2021}. No obvious QPOs were observed in the power spectrum. It should be noted, that low frequency QPOs were previously observed in this system during short episodes of spectral hardening within the soft state \citep{1989PASJ...41..519E, Alam2014}.

\begin{figure} 
\centering
\includegraphics[width=7.8cm]{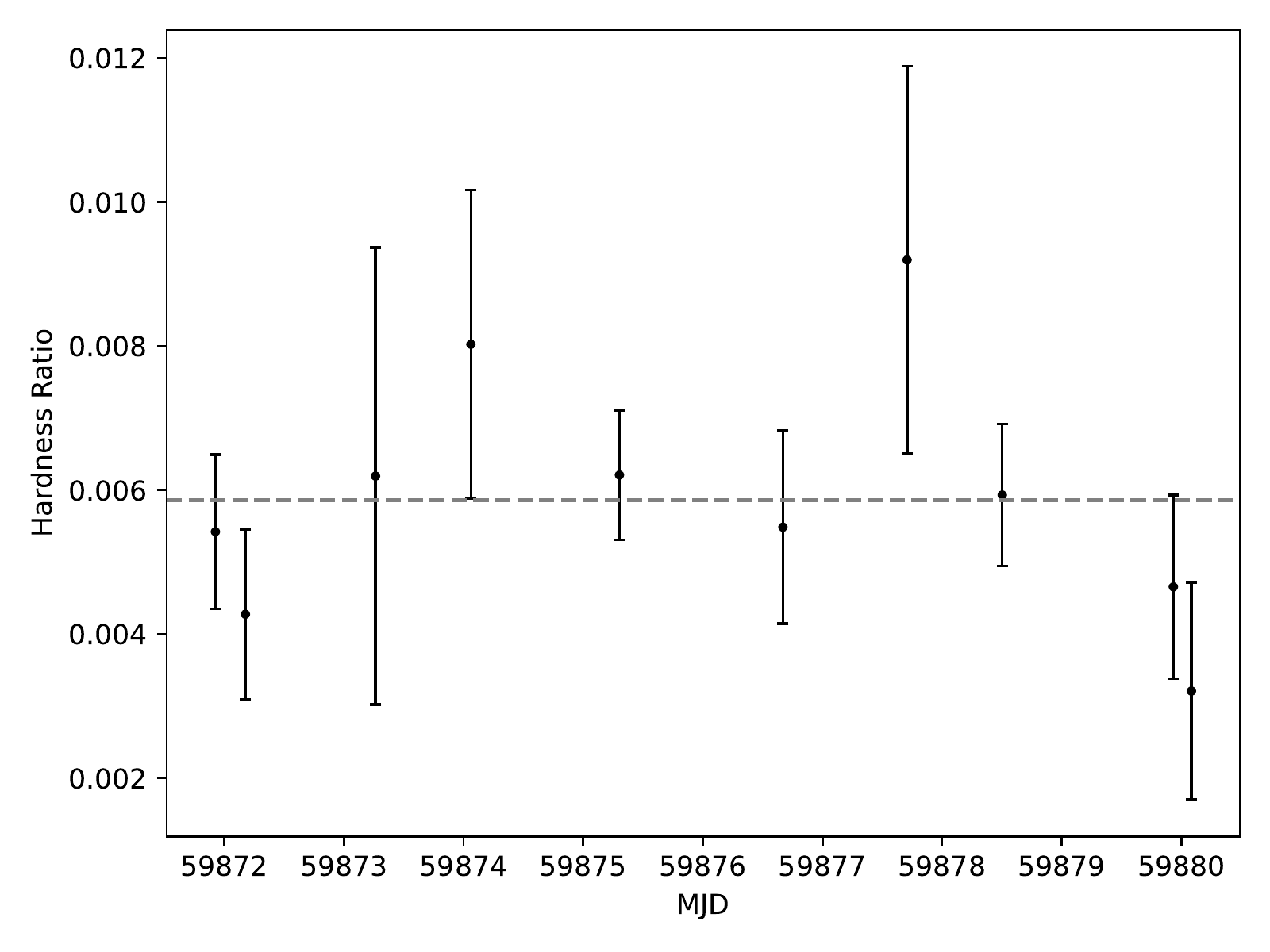} 
\includegraphics[width=8cm]{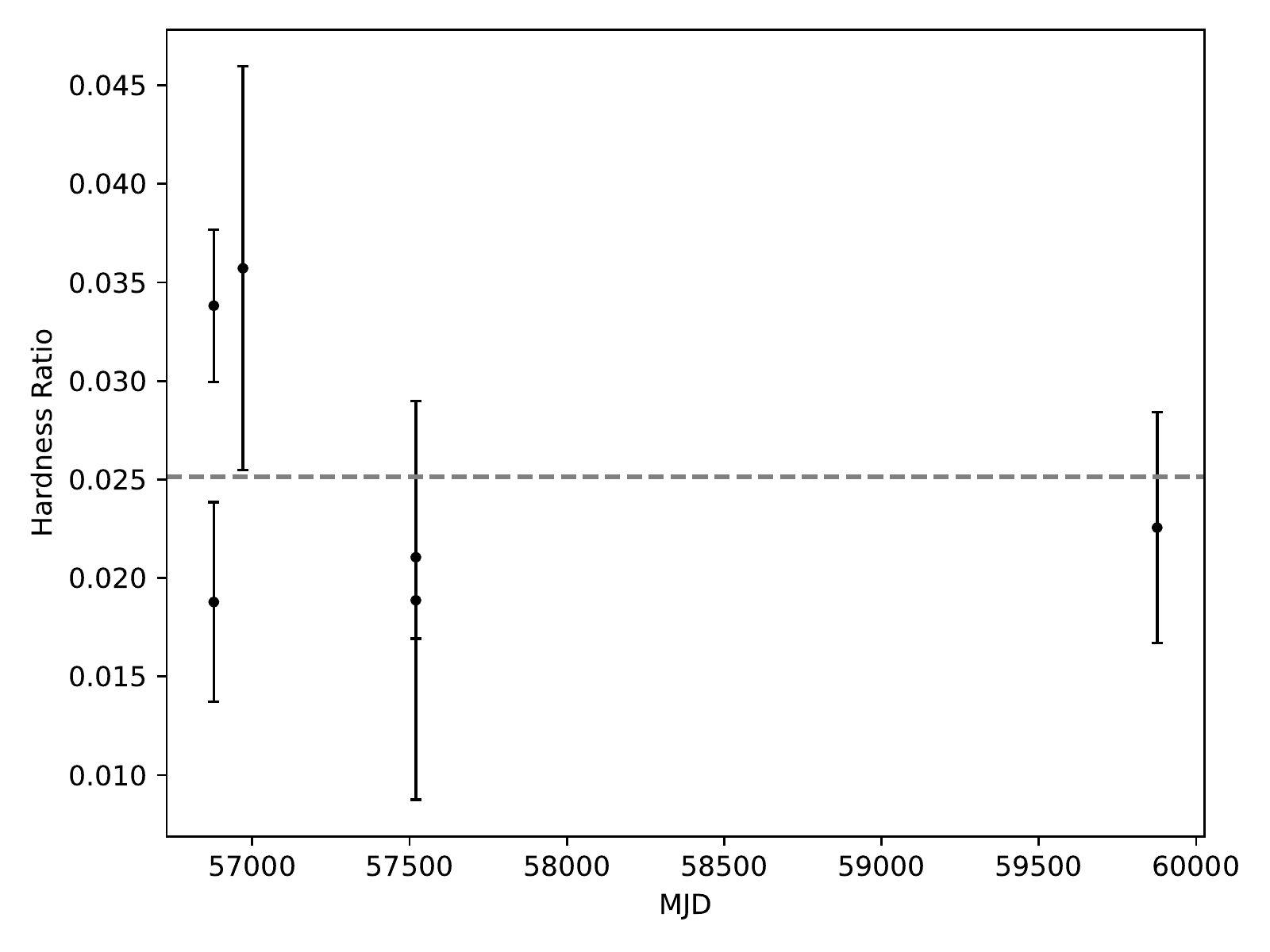}
\caption{Time variation of X-ray hardness ratios. The dashed lines are the average values of the hardness ratios. \textit{Top panel}: Ratio of the {NICER} count rates in the hard band (3--12 keV) over the total flux (0.3--12 keV). \textit{Bottom panel}: Ratio of the {\em NuSTAR} count rates in the hard band (8--20 keV) over the total flux (3--20 keV).}
\label{fig:HR}
\end{figure}

Using the {NICER} and {\em NuSTAR} spectral data, we calculated the hardness ratio defined as the ratio between the flux in the hard band and the total flux. We defined the soft vs. hard bands to be 0.3--3 keV vs. 3--12 keV for {NICER}, and 3--8 keV vs. 8--20 keV for {\em NuSTAR}. In Fig. {\ref{fig:HR}}, we show the evolution of the hardness ratio for the {NICER} and {\em NuSTAR} data. The {NICER} hardness ratio is consistent with being constant with an average hardness of 0.0059 within the measurement uncertainties.
The low hardness indicates that the source is in the soft state when the accretion-disc thermal emission clearly dominates in the X-ray spectrum.
The average {\em NuSTAR} hardness ratio is 0.025 for the simultaneous observation with {\em IXPE}.

Because the spectral hardness appears constant over the observations, and given the stability in flux, we used {NICER}, {\em NuSTAR}, and {\em IXPE} time-averaged spectra for the spectral fitting. We fit each of the 10 {NICER}, 2 {\em NuSTAR}, and 3 {\em IXPE} spectra individiually, as a joint fit.
We used the {\em NuSTAR} spectra up to $20$ keV as the background signal becomes comparable to that of the source at higher energies. The {ART-XC} data were not used for detailed spectral analysis, because of too high noise.
We used the \textsc{xspec} package
and employed the following Model \ref{eq:specmodel} for the time-averaged analysis:
\vspace{-1mm}
\begin{equation} \label{eq:specmodel}
{\tt GABS \times TBFEO \ (GAUSSIAN + KERRBB + NTHCOMP).}\\
\end{equation}

We used {\tt KERRBB} \citep{Li2005} to model general relativistic accretion disc emission from a multi-temperature blackbody and {\tt NTHCOMP} \citep{Zdziarski1996, Zycki1999} for the thermally Comptonized continuum. For the {\tt KERRBB} model, we kept the  BH mass and distance fixed at the values reported for the source ($M_{\rm BH}=10.9 \ \mathrm{M_\odot}$, $d = 50$ kpc) and assumed the disc axis to be aligned with the binary system orbital inclination ($i=36.4\degr$), i.e. the disc is not warped. We fixed the dimensionless spin parameter of the BH to the best-fitting value of $0.92$ found with the continuum fitting method by \cite{Gou2009}. We also kept the spectral hardening factor fixed at $1.7$, and assumed no torque at the inner disc edge.

The blackbody seed photon temperature $kT_\textrm{bb}$ of the {\tt NTHCOMP} model was allowed to vary in the range 0.4--1.0 keV. The lower limit
was obtained from prior modeling where $kT_\textrm{bb}$ was tied to the $kT_\textrm{in}$ of the multi-blackbody model {\tt DISKBB} to calculate the temperature of the inner edge of the accretion disc and the Compton up-scattering of seed photons at this temperature. The upper limit is set to the maximum $kT_\textrm{in}$ fitted to archival data reported in \citet{Gou2009}. The blackbody seed photon temperature was $0.888 \pm 0.005$ keV,  consistent with values reported in \citet{2001MNRAS.325.1253G} and \citet{Kubota2005}.
We find a photon index of $2.60 \pm 0.02$, well within previously reported ranges employing the {\tt NTHCOMP} \citep{Jana2021} and {\tt POWERLAW} \citep{Jana2021} models.
The \textit{int\textunderscore type} parameter of {\tt NTHCOMP} is set to $0$ for blackbody seed photons.

A {\tt GAUSSIAN} component was added at $0.88$ keV with a line width of $0.25$ keV to account for an emission feature that resembles the first-order scattering of anisotropic photons onto isotropic electrons like that in \cite{Zhang2019}, fig. 8.  The observation 3 of {NICER} presented a more pronounced {\tt GAUSSIAN} component that required different line energy and normalization parameter values with the line width consistent to other {NICER} observations within the 90 percent confidence interval. {\tt GABS} was used to model a broad Gaussian-like absorption artifact at 9.66 keV detected with {\em NuSTAR} that may be due to Comptonization in the upper layers of the disc not being modeled properly, an inhomogeneous corona, a broad instrumental absorption feature, or an unmodelled weak reflection component. The line energies for both of the identified emission and absorption-like features, $E_\textrm{l}$ in {\tt GAUSSIAN} and {\tt GABS} respectively, are left frozen while their line widths and normalization/depth are allowed to vary freely.

{\tt TBFEO} \citep{Wilms2000} was used to account for the X-ray absorption by hydrogen, oxygen, and iron. The fitted equivalent hydrogen column which accounts for absorption in our Galaxy, in the Large Magellanic Cloud and in the binary system was $(0.938\pm 0.001) \times 10^{22}$\,cm$^{-2}$. We note that while this value is smaller than the $(1.0 - 1.3) \times 10^{22}$\,cm$^{-2}$ reported in \citet{Hanke2010}, these higher values worsen the fit. Although the metallicity should vary along the line of sight, we use a single absorber for simplicity. The iron and oxygen abundances relative to Solar are allowed to vary freely.

\begin{figure}
    \centering
    \includegraphics[width=8.5cm]{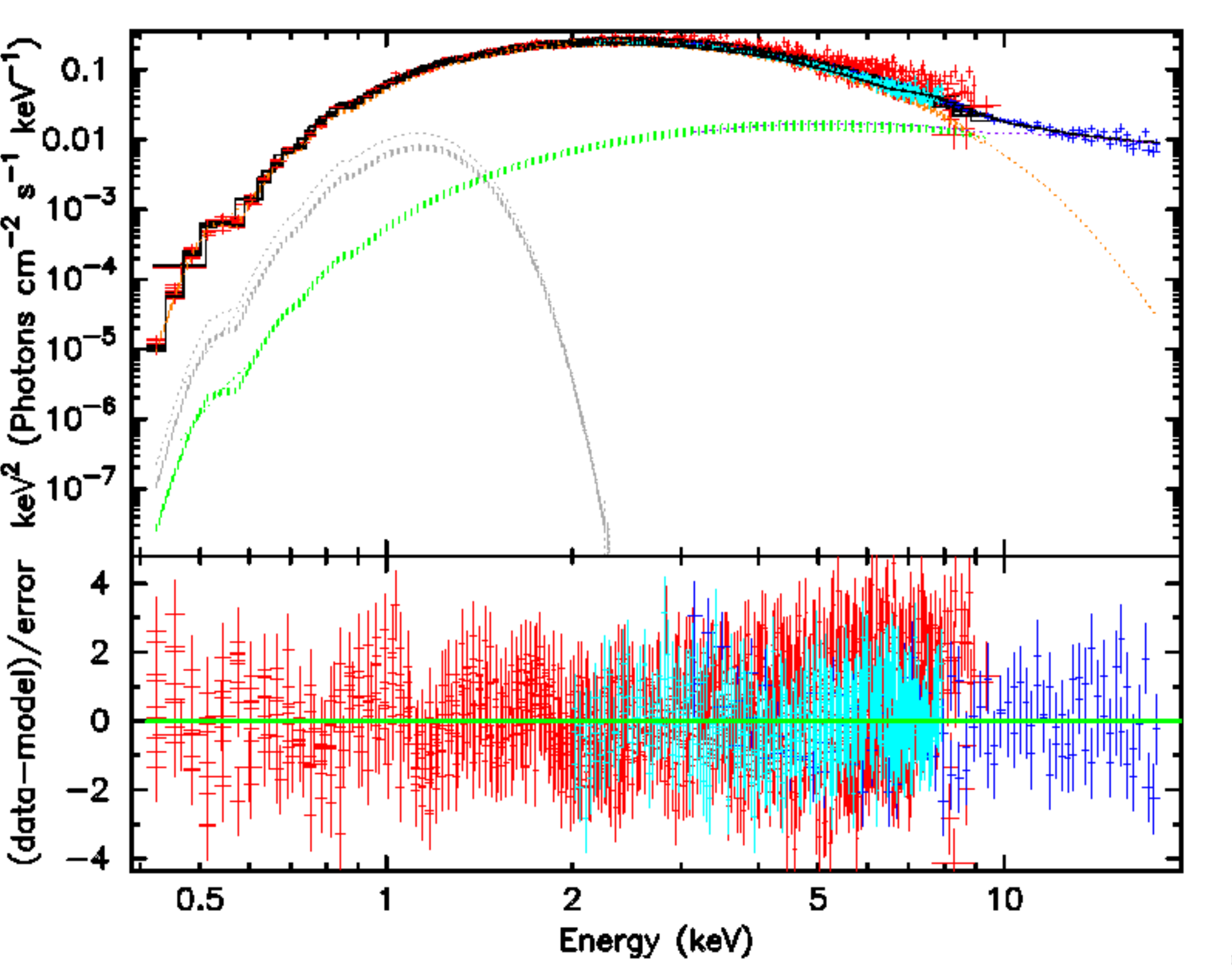}
    \caption{X-ray spectra of LMC X-1. \textit{Top panel}: {NICER} (red), {\em NuSTAR} (blue), and {\em IXPE} (cyan) spectra unfolded around the best-fitting model described by Model \ref{eq:specmodel} in $E F(E)$ space. The total model for each data set is shown in black with individual {\tt GAUSSIAN}, {\tt KERRBB}, and {\tt NTHCOMP} contributions in light gray, orange, and green, respectively. \textit{Bottom panel}: Model-data deviations (residuals) in $\sigma$.}
    \label{fig:spectralfit}
\end{figure}

We find the best-fitting model has $\chi^2 / {\rm dof} = 3497.83/2571$. We estimate a BH accretion rate of $\Dot{M}=(1.756 \pm 0.002)\times 10^{18}$\,g\,s$^{-1}$, consistent with values reported for the source in \citet{Zdziarski2023}.
The flux in the 2--8 keV energy range is dominated by the accretion disc emission with {\tt KERRBB} contributing 94 per cent, while the coronal emission ({\tt NTHCOMP}) contributes 6 per cent. Figure \ref{fig:spectralfit} shows the unfolded spectra and the best-fitting parameters as reported in Table \ref{t:spectralfitparams}.

\renewcommand{\thetable}{1}
\begin{table}
\begin{center}
\resizebox{\columnwidth}{!}{
\begin{tabular}{p{1.5cm}p{2cm}p{2.5cm}p{2cm}}
\hline
\hline
Component & Parameter (unit) & Description & Value\\  
\hline
\hline
{\tt TBFEO} & $N_\textrm{H}$ ($10^{22}$\,cm$^{-2}$) & Hydrogen column density & $0.938\substack{+0.001 \\ -0.001}$ \\
& O & Oxygen abundance & $0.882\substack{+0.004 \\ -0.004}$\\
& Fe & Iron abundance & $0.78\substack{+0.01\\ -0.01}$\\
& $z$ & Redshift & $0.0$ (frozen)\\
\hline
{\tt KERRBB} & $\eta$ & Inner-edge torque & $0.0$ (frozen) \\
    & $a$ & Black-hole spin & $0.92$ (frozen) \\
    & $i$ (deg) & Inclination & $36.4$ (frozen) \\
    & $M_{\textrm{bh}}$ (M$_\odot$) & Black-hole mass & $10.9$ (frozen) \\
    & $M_{\textrm{dd}}$ ($10^{18}$\,g s$^{-1}$) & Mass accretion rate & $1.756 \substack{+0.002 \\ -0.002}$ \\
    & $D_\textrm{bh}$ (kpc) & Distance & $50$ (frozen) \\
    & $hd$ & Hardening factor & $1.7$ (frozen) \\
    & $r_\textrm{flag}$ & Self-irradiation & $1$ (frozen) \\
    & $l_\textrm{flag}$ & Limb-darkening & $0$ (frozen) \\
    & norm & Normalization & $1.0$ (frozen) \\
\hline
{\tt NTHCOMP} & $\Gamma$ & Photon index &
$2.60\substack{+0.02\\-0.02}$\\
& $kT_\textrm{e}$ (keV) & Electron temperature &
$100.00$ (frozen)\\
& $kT_\textrm{bb}$ (keV) & Seed photon temperature & $0.888\substack{+0.005\\-0.005}$\\
& norm (10$^{-3}$) & Normalization & $2.23\substack{+0.03\\-0.03}$\\
\hline
{\tt GAUSSIAN} & $E_\textrm{l}$ (keV) & Line energy & $0.88$ (frozen)\\
& $\sigma$ (keV) & Line width & $0.25\substack{+0.01\\-0.01}$\\
& norm (10$^{-2}$ photons cm$^{-2}$ s$^{-1}$) & Normalization & $1.74\substack{+0.06\\ -0.06}$\\
\hline
{\tt GABS} & $E_\textrm{l}$ (keV) & Line energy & $9.66$ (frozen) \\
& $\sigma$ (keV) & Line width & $0.9\substack{+0.2 \\-0.2}$\\
& Strength (keV) & Line depth & $0.22\substack{+0.06\\-0.06}$\\
\hline
\end{tabular}
}
\caption{Best-fitting parameters (with uncertainties at 90 per cent confidence level) of the joint {NICER}, {\em NuSTAR}, and {\em IXPE} spectral modeling with the combined model described by Model \ref{eq:specmodel}. $\chi^2 / {\rm dof}$ for the fit is $\chi^2 / {\rm dof} = 3497.83/2571$. {\tt GAUSSIAN} parameter values for the {NICER} observation 3 are: $E_\textrm{l} = 0.85$ keV and norm $= 0.034 \pm  0.001$ photons cm$^{-2}$ s$^{-1}$. See Appendix \ref{observations_remaining} for discussion of the normalization of the {\tt KERRBB} component.}
\label{t:spectralfitparams}
\end{center}
\end{table}

The obtained $\chi^2 / {\rm dof}$ for the best-fitting model is greater than 1, despite the addition of systematic errors (see Appendix \ref{observations_remaining}).
This may be due to several reasons: cross-calibration uncertainties between the different instruments, short term source variability, different exposure intervals of the various satellites, and complexity of the X-ray spectra of Galactic BHs which may be not fully captured by the model. However, as a detailed spectral analysis is beyond the scope of the paper and a visual inspection of the residuals seems to indicate that the global fit is not obviously incorrect, we used the best-fitting model to derive the polarization properties of the various spectral components.

\subsection{Polarimetric analysis with PCUBEs}\label{pol_analysis}

\begin{figure}
    \centering
    \includegraphics[width=8.5cm]{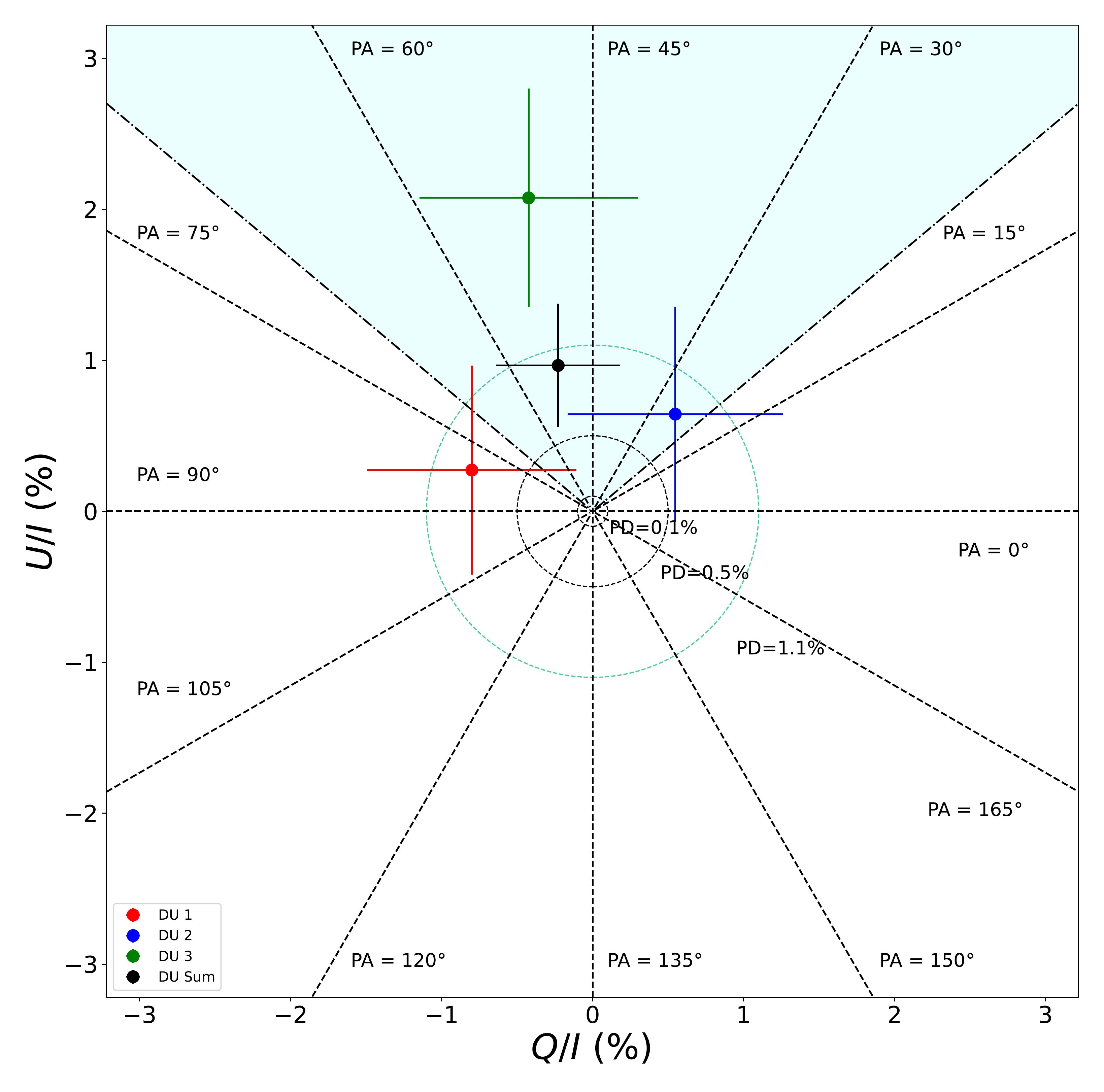}
    \caption{Normalized $Q/I$ and $U/I$ Stokes parameters and corresponding polarization degree and angle for DU1 (red), DU2 (green), DU3 (blue) and a sum of the three units (black). The (light green) circle represents the MDP value at the 99 per cent confidence level and the cyan-shaded area the direction and projected full opening angle of the ionization cone. The data are obtained using a single energy bin in the 2--8 keV energy band. We report the uncertainties at 1$ \sigma $ level, i.e. at the 68.3 per cent confidence.}
    \label{fig:pcube}
\end{figure}

We show in Fig. \ref{fig:pcube}  the normalized Stokes parameters ($Q/I$ and $U/I$) for a single energy bin 2--8 keV, for each DU separately, and summed. The polarization angle measured by {\em IXPE} using the sum of all three DUs is $51.6\degr \pm 11.8\degr$ in the north-east direction and the polarization degree is $1.0 \pm 0.4$ per cent. Given this measurement, we have a $3\sigma$ upper limit on polarization degree of 2.2 per cent. The polarization angle value is roughly aligned with the ionization cone structure detected in He II $\lambda 4686/ \textrm{H}\beta$ and [O III] $\lambda 5007/ \textrm{H}\beta$ line ratio maps at 225\degr\ north-east (with a projected full opening angle 50\degr) \citep{Cooke2007, Cooke2008}, because the polarization angle is defined modulo 180\degr. The obtained low value of the upper limit on the polarization degree is consistent with the classical results of \citet{Chandrasekhar1960} and \citet{Sob63}, approximated by equation (41) of \citet{Viironen2004}, for scattering-induced polarization of pure thermal emission in semi-infinite disc atmospheres seen at inclination below $\sim 60\degr$. However, see Section \ref{section:Spectro_pol_analysis} for a careful discussion of the polarization result with respect to the two observed spectral components. The MDP at 99 per cent confidence level in 2--8 keV is 1.1 per cent, which means the obtained polarization result is not statistically significant. Reducing the energy range does not improve the statistical significance.

\begin{figure}
    \centering
    \includegraphics[width=8.5cm]{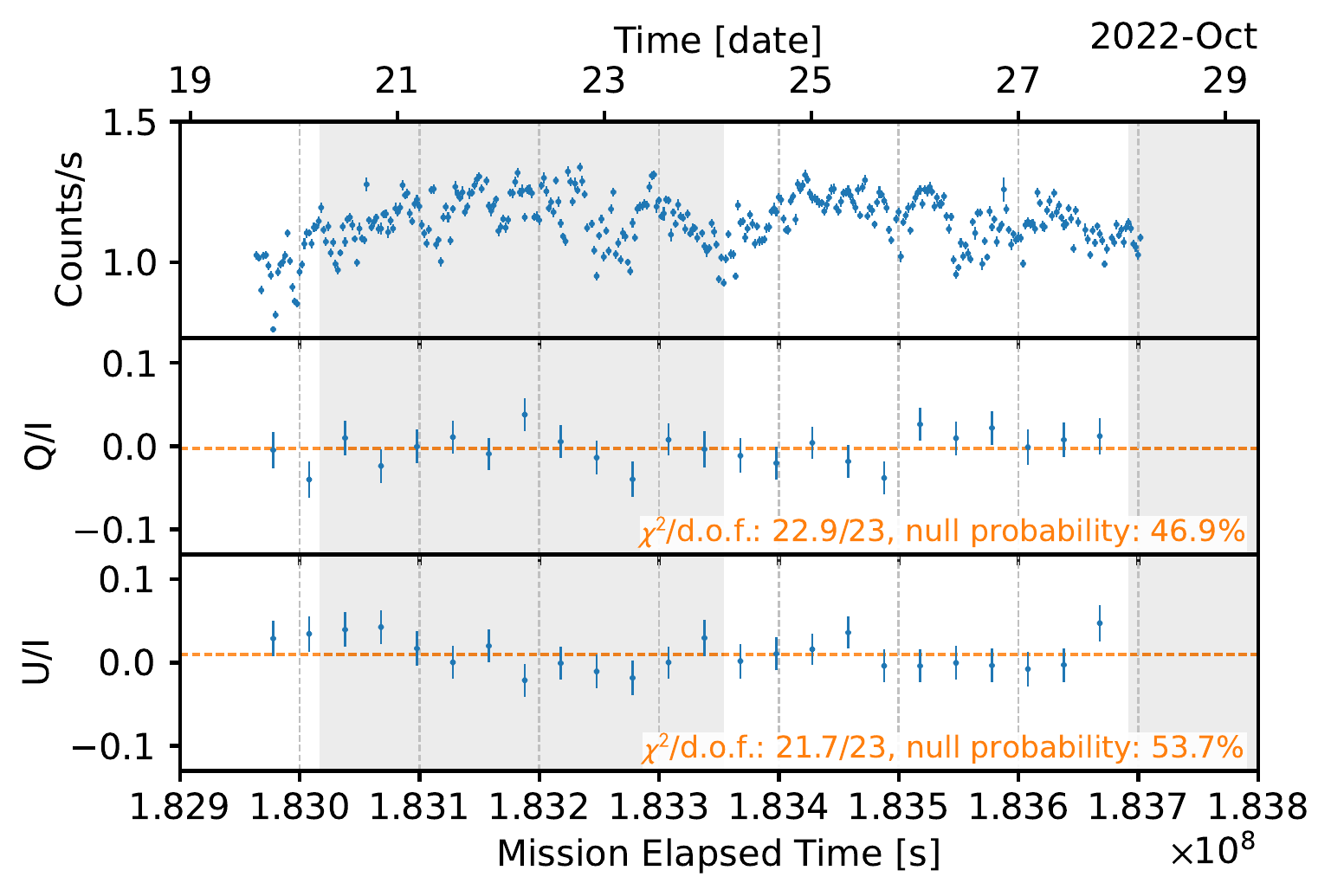}
    \caption{Counting rate ({\em top}) and normalized Stokes $Q$ and $U$ parameters ({\em middle} and {\em bottom}, respectively) measured by {\em IXPE} as a function of time. Time bin is 2~ks for counting rate and 30~ks for $Q$ and $U$. Horizontal, dashed lines are the best fit with a constant line: the obtained $\chi^2$, the number of degrees of freedom and the corresponding null probability are indicated. The gray-shaded and white regions identify subsequent orbits of LMC~X-1.}
    \label{fig:stokes_time}
\end{figure}

Although no average polarization is observed, a \emph{time-dependent} signal may still be present in the {\em IXPE} observation. To check for this possibility, we adopted the dedicated \textsc{ixpeobssim} function to calculate the normalized Stokes parameters $Q/I$ and $U/I$ in time bins of 30~ks (see Fig.~\ref{fig:stokes_time}). These can be considered independent normal variables \citep{Kislat2015} and we fit their values as a function of time with a constant line. The fit null probability, which expresses the probability that the observed variations around the model are due to chance alone, is $\approx 50$ per cent for both $Q/I$ and $U/I$ for the value of the $\chi^2$ found and the number of degrees of freedom of the fit. Then, we derived that any observed variability of polarization is compatible with statistical fluctuations only.

We repeated a similar procedure to investigate possible dependence of polarization on the orbital phase. We first derived the phase of each event from its arrival time using the orbital ephemeris described in Section \ref{spectral_analysis}. Then, we folded the events into 7 phase bins. Variations of the normalized Stokes parameters in the entire 2--8~keV {\em IXPE} energy band are compatible with statistical fluctuations: summing the $\chi^2$ values obtained for the fit of both the Stokes parameters and, correspondingly, their degrees of freedom, the null probability of the combined fit is 1.1 per cent. However, selecting only the events in the 2--4~keV energy range, the null probability is reduced to 0.0057 per cent.
This further supports the fact that the emission from LMC~X-1 may indeed be polarized at a few per cent, but its polarization angle, degree, or both, could depend on the orbital phase. When summing over time scales comparable to the orbital period, an orbital-phase-dependent polarization would be averaged to a low value that would be undetected in the phase-average analysis. However, {\em IXPE} observed only two complete orbits of LMC~X-1 (see Fig.~\ref{fig:stokes_time}); therefore further observations would be needed to detect orbital-phase-dependent polarization with high statistical confidence.

\begin{figure}
    \centering
    \includegraphics[width=8.5cm]{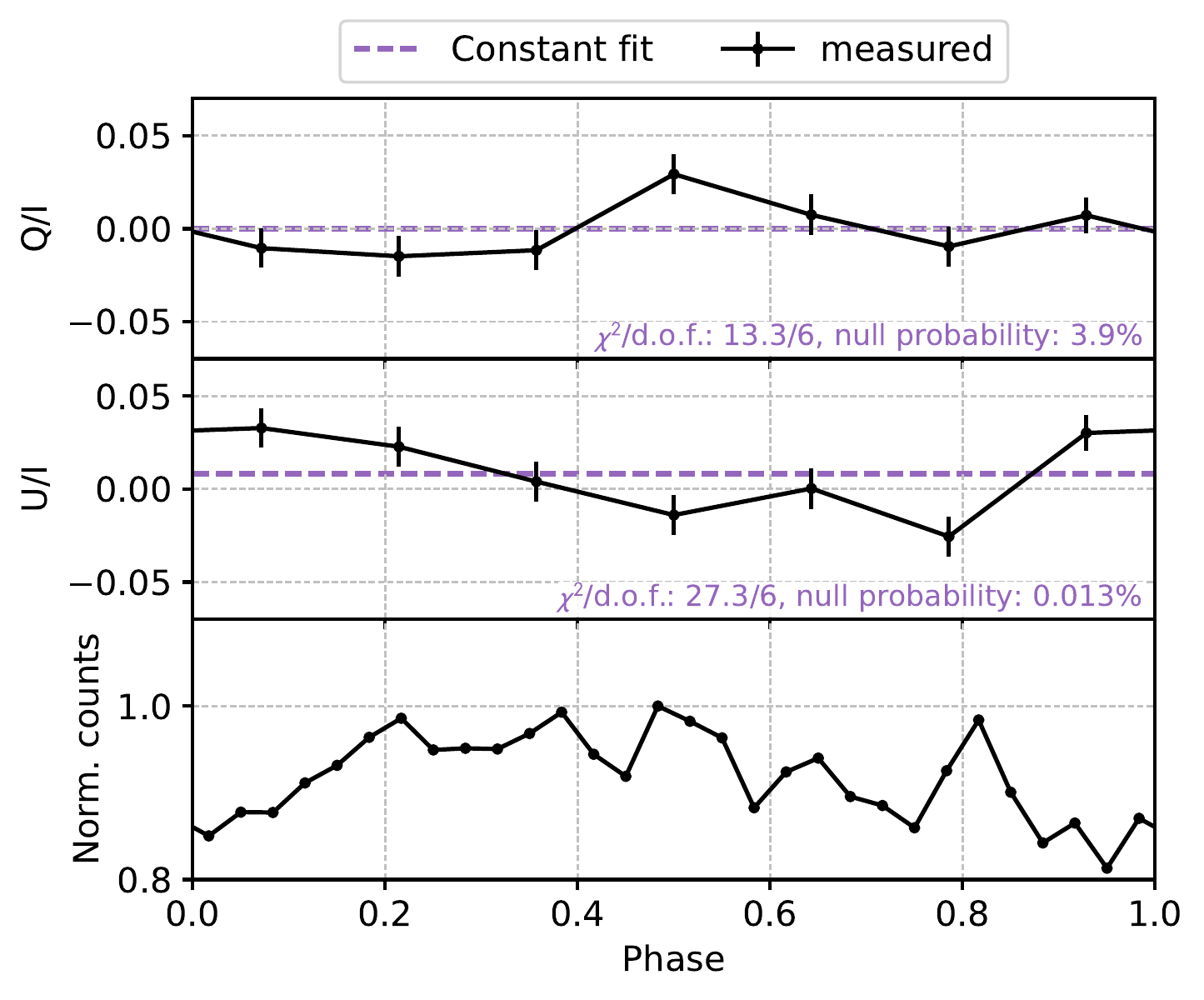}
    \caption{Variation of the normalized Stokes parameters $Q$ ({\em top}) and $U$ ({\em middle}), calculated with \textsc{ixpeobssim}, as a function of the orbital phase of LMC~X-1. As in Fig.~\ref{fig:stokes_time}, horizontal, dashed lines are the best fit with a constant line, and the $\chi^2$, the number of degrees of freedom and the corresponding null probability of the fit are indicated. The corresponding flux from {\em IXPE}, normalized to its maximum observed value, ({\em bottom}) as a function of the orbital phase is added for comparison.}
    \label{fig:stokes_orbit}
\end{figure}

\subsection{Polarimetric analysis with \textsc{xspec}}
\label{section:Spectro_pol_analysis}

For the polarimetric fit of our data, we removed the {NICER} and {\em NuSTAR} spectra and included the {\em IXPE} $Q$ and $U$ spectra. Since our aim here was to explore the polarimetric properties of the source with the simplest possible model, we removed both $\tt GAUSS$ and $\tt GABS$ component from Model \ref{eq:specmodel} and we convolved the thermal and the Comptonized components with the polarization model $\tt POLCONST$; this is characterized by two parameters, the polarization degree $\Pi$ and angle $\Psi$, both constant with energy. Thus we employed Model \ref{eq:Model2} in the fitting procedure defined as follows:
\begin{equation}
    \tt TBFEO * (POLCONST*KERRBB+POLCONST*NTHCOMP) .
    \label{eq:Model2}
\end{equation}
We maintained the spectral parameters frozen at the values shown in Table \ref{t:spectralfitparams}, while allowing both components' polarization degree and angle to vary freely during the fitting procedure. As a result, we obtained a best-fit $\chi^2 /{\rm dof} = 842.5/894$, with the polarization parameters values listed in Table \ref{t:spectropolfitparams}. 

\begin{figure}
    \centering
    \vspace{-15pt}
    \subfloat{\includegraphics[width=8.5cm]{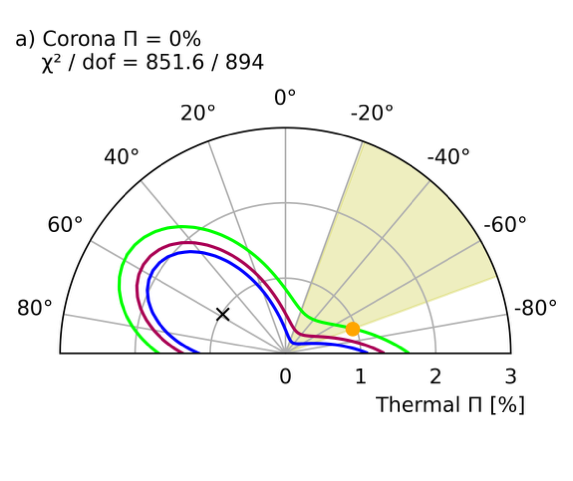}}\\
    \vspace{-40pt}
    \subfloat{\includegraphics[width=8.5cm]{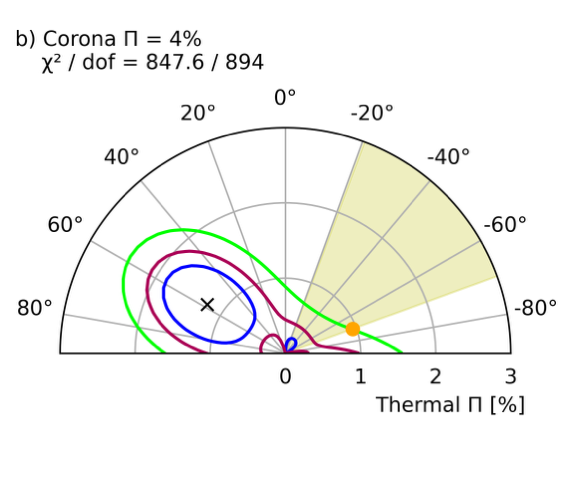}}\\
    \vspace{-40pt}
    \subfloat{\includegraphics[width=8.5cm]{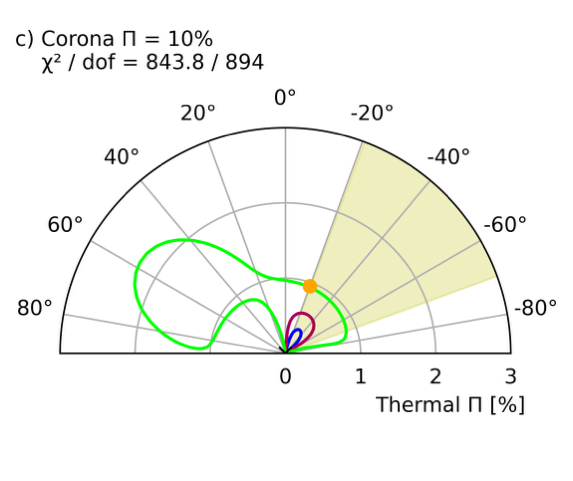}}\\
    \vspace{-30pt}
    
    \caption{Contour plots of the polarization degree $\Pi$ and angle $\Psi$ associated to the accretion disc thermal emission. Blue, red, and green lines indicate 68, 90, and 99 per cent confidence levels for two parameters of interest, respectively. The black cross indicates the best-fit parameters for the $\chi^2/{\rm dof}$ value shown in the label. The coronal emission is assumed to be polarized perpendicularly to the thermal component, and its polarization degree is fixed at 0 (\emph{top}), 4 (\emph{middle}), and 10 per cent (\emph{bottom}). The yellow-shaded region indicates the projected accretion disc plane, perpendicular to the projected ionization cone. The orange dots represent the 3$\sigma$ upper limit of thermal emission polarization degree, assuming that this component is polarized in the same direction as the projected accretion disc plane, i.e. perpendicularly to the observed projected ionization cone direction. The accretion disc is assumed to be aligned with the orbital inclination $i=36.4\degr$.}
    \label{fig:CPlots_PD_PA}
\end{figure}

Because we obtained only an upper limit on the polarization degree, we were not able to constrain the polarization properties of both spectral components at the same time. Thus we decided to further analyze the polarimetric data by tying the two components' polarization angles. In particular the polarization degree and angle associated with the accretion disc thermal emission were left free to vary, while we linked the polarization angle of the coronal emission to that of the thermal emission. Because of the symmetry of the system, the polarization vector of the thermal emission is expected to be either parallel or perpendicular to the disc symmetry axis. However, the Chandrasekhar-Sobolev result and many simulation studies suggest that the thermal emission is locally likely to be polarized perpendicular to the disc symmetry axis. This is especially true when considering optically-thick disc atmospheres with large optical depth \citep{Dovciak2008,Taverna2020}, or when accounting for absorption processes alongside scattering ones \citep{Taverna2021}. The coronal emission polarization vector can be either parallel or perpendicular to the disc axis, depending on the corona geometry, its location, and velocity \citep[see e.g.][]{Zhang2022}. Nevertheless, the recent observation of Cyg X-1 \citep{Krawczynski2022} as well as theoretical predictions for a flat corona sandwiching the disc \citep[see e.g.][]{PS1996,Schnittman2010, Krawczynski2022b} suggest that this component is polarized in the same direction as the disc axis. Hence, we forced the polarization vectors of the two components to be perpendicular to each other. In this configuration, the total polarization degree of the model is given by the difference between the two components' contribution, effectively allowing for two unphysically large polarization degree values at the same time. To avoid this, we restricted our analysis to three reasonable values for the coronal emission polarization degree: 0, 4  \citep[the best-fitting value for coronal emission polarization degree found for Cyg X-1 in][]{Krawczynski2022}, and 10 per cent. The resulting contour plots for the polarization degree and angle of the thermal emission are shown in Fig. \ref{fig:CPlots_PD_PA}. The ionization cone orientation of $\sim$ 45\degr\ suggests that the projected accretion disc plane is perpendicular to the projected jet-remnant direction \citep[see e.g.][]{Krawczynski2022}, i.e. approximately $-45\degr \pm 25\degr$ in our plots, which is marked by the yellow-shaded region in Fig. \ref{fig:CPlots_PD_PA}, taking into account the observed projected full opening angle of the ionization cone. Thus the thermal component is expected to be polarized in this direction.

\renewcommand{\thetable}{2}
\begin{table}
\begin{center}
\resizebox{\columnwidth}{!}{
\begin{tabular}{p{1.5cm}p{2cm}p{2.5cm}p{2cm}}
\hline
\hline
Component & Parameter (unit) & Description & Value\\  
\hline
{\tt POLCONST} & $\Pi$ ($\%$) & Polarization degree & $\leq 1.6$\\
{\tt (1)} & $\Psi$ (deg) & Polarization angle & Unconstrained\\
\hline
{\tt POLCONST} & $\Pi$ ($\%$) & Polarization degree & $\leq 35.3$ \\
{\tt (2)} & $\Psi$ (deg) & Polarization angle & Unconstrained\\
\hline
\end{tabular}
}
\caption{Best-fit parameters of the {\em IXPE} polarimetric analysis described in Section \ref{section:Spectro_pol_analysis}. The components {\tt POLCONST (1)} and {\tt (2)} are used in Model \ref{eq:Model2} to describe the polarization properties of the disc and the corona emission, respectively.}
\label{t:spectropolfitparams}
\end{center}
\end{table}

When assuming the coronal emission to be unpolarized, we found an upper limit of 2.5 per cent on the thermal emission polarization degree, while forcing the polarization angle to be directed as the projected accretion plane this value reduces to 1.0 per cent, which is marked by the orange dot in top panel of Fig. \ref{fig:CPlots_PD_PA}. When taking into account the coronal emission polarization, the contour plots show two minima, representing two allowed configurations. In one case the thermal component is polarized in the same direction as the projected accretion plane with a low polarization degree, while in the other it is polarized perpendicularly to it, but with a larger polarization degree. In both cases, the polarization degree upper limits tend to increase, becoming as high as $\Pi=2.4$ and $\Pi=2.2$ per cent when the Comptonized component polarization degree is fixed at 4 and 10 per cent, respectively; and $\Pi=0.9$ and $\Pi=0.9$ per cent, if we further assume the suggested system orientation, which is marked by the orange dots in middle and bottom panels of Fig. \ref{fig:CPlots_PD_PA}. These polarization degree values are all well within the Chandrasekhar estimates for the polarization of thermal radiation. The polarization angle value is unconstrained at the 99 per cent confidence level in all cases (see Fig. \ref{fig:CPlots_PD_PA}).

We also attempted for a joint spectro-polarimetric fit in \textsc{xspec}, using a physical model of thermal emission {\tt KYNBBRR} \citep{Taverna2020,Mikusincova2023}, while keeping the phenomenological constant polarization prescription to the power-law component. The {\tt KYNBBRR} model is an extension of the relativistic package KYN \citep{Dovciak2004,Dovciak2008}, developed to include the contribution of returning radiation, i.e. photons that are bent by strong gravity effects and forced to return to the disc surface, where they can be reflected before eventually reaching the observer \citep{Schnittman2009,Taverna2020}. Although this approach in theory allows to put constraints on the polarization of the Comptonization component, the BH spin and the accretion disc inclination, we could not obtain any reasonable restrictions on these parameters, given our spectral and polarization data.

\subsection{Physical polarization model of Comptonized and thermal emission}\label{compps_modelling}

\begin{figure}
    \centering
    \includegraphics[width=8.5cm]{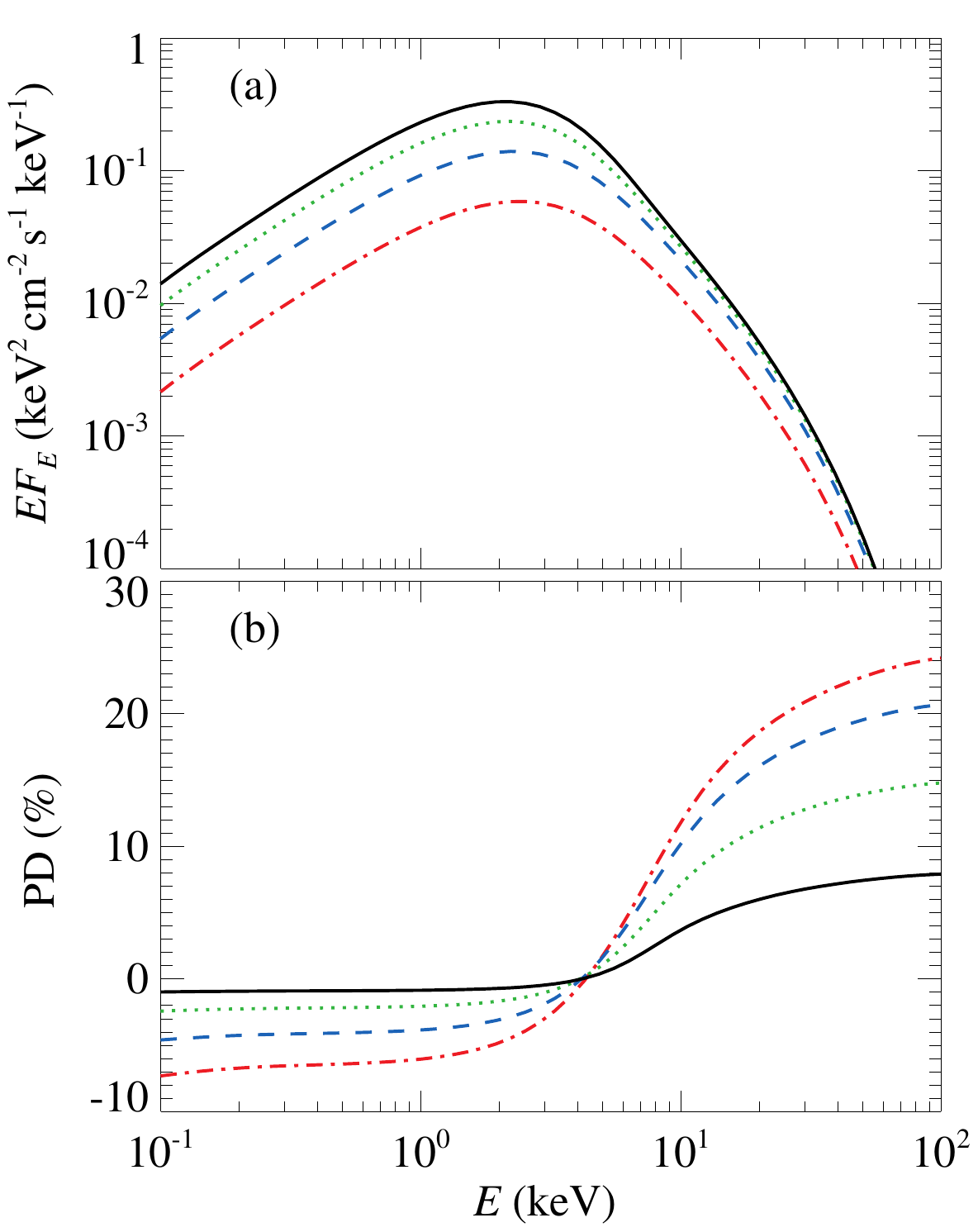}
    \caption{Spectral energy distribution (a) and polarization degree (b) obtained for the slab corona model. Lines correspond to different inclinations: $i=30\degr$ (black solid), $45\degr$ (green dotted), $60\degr$ (blue dashed) and $75\degr$ (red dot-dashed).}
    \label{fig:slab_corona}
\end{figure}

In addition, we performed simulations of a slab coronal geometry with a cold disc and a hot Comptonization medium above it using a radiative transfer code that splits the radiation field produced by Compton scattering in different orders and computes their intensities, source functions and polarization \citep{veledina22,Poutanen2023}.
The code follows the procedures described in \citet{PS1996}. For consistency with the spectral data, we performed additional spectral fit with this model that is described in Appendix \ref{compps_fit}. The obtained values served as referential for the polarization modeling. We assumed the slab is illuminated by the accretion disc whose radiation is described by the multi-temperature blackbody $kT_{\rm bb}=0.81$~keV and angular distribution and polarization follow the Chandrasekhar-Sobolev profile \citep{Chandrasekhar1960,Sob63}.
The temperature of the medium is assumed to be $kT_{\rm e}=10$~keV and the Thomson optical depth to be $\tau_{\rm T}=1.26$.
We plot the resulting spectra for different inclinations ($i=30\degr, 45\degr, 60\degr$ and $75\degr$) in Fig.~\ref{fig:slab_corona}a.
In Fig.~\ref{fig:slab_corona}b we show the polarization degree corresponding to this geometry.
Positive(/negative) values correspond to polarization parallel(/orthogonal) to the disc axis.

The change of polarization sign at $\sim5$~keV is a known feature of the slab corona geometry \citep[see e.g.][]{PS1996}, as the sign of each Compton scattering order is controlled by the angular distribution of the incoming (seed) photons.
We find that, for the considered parameters, the switch between negative and positive polarization degree occurs in the middle of {\em IXPE} range.
This might be the reason for the low net polarization degree averaged over the entire 2--8~keV band, and can plausibly serve as a mechanism for switching between the positive and negative polarization degrees seen in Fig.~\ref{fig:stokes_orbit}: variations of the parameters lead to variations of the characteristic energy of zero polarization.
In this case, the variations likely have a stochastic, rather than periodic (e.g. at orbital period) origin.

\section{Conclusions}\label{conclusions}

We performed a broadband X-ray spectro-polarimetric observational campaign of the BH binary system LMC X-1 simultaneously with the {\em IXPE}, {NICER}, {\em NuSTAR} and {ART-XC} missions. The spectral data are consistent with previous studies of LMC X-1. We report that the source is in the high/soft state with a dominant thermal component in the X-ray band, a power-law Comptonization component that begins to prevail around $\sim 10$ keV, and a negligible reflection contribution. The spectra do not show significant time variability. The first X-ray polarimetric observation of LMC X-1 by {\em IXPE} constrains the polarization degree to be below the MDP of $1.1$ per cent at the 99 per cent confidence level for the time-averaged emission in the 2--8 keV band. This is consistent with theoretical predictions for pure thermal emission from a geometrically thin and optically thick disc with a Novikov-Thorne profile, assuming Chandrasekhar's prescription for polarization due to scattering in semi-infinite atmospheres. Spectro-polarimetric fitting leads to upper limit (at 99 per cent confidence level) on the polarization degree of the thermal radiation to be 1.0, 0.9 or 0.9 per cent when the polarization of power-law component is fixed to 0, 4 or 10 per cent, respectively, if the two components are polarized perpendicular to each other and if we assume a preferred system orientation given by the optical data from literature. The new X-ray polarimetric data show hints of non-zero polarization with the polarization angle aligned with the projected ionization cone and weak evidence for time variability of the polarization that could be attributed to a stochastic origin in a slab corona scenario sandwiching a thermally radiating accretion disc. The 562 ks observation by {\em IXPE} did not allow statistically significant constraints on the BH spin nor the disc inclination.

\section*{Acknowledgements}

The \textit{Imaging X-ray Polarimetry Explorer} ({\em IXPE}) is a joint US and Italian mission.  The US contribution is supported by the National Aeronautics and Space Administration (NASA) and led and managed by its Marshall Space Flight Center (MSFC), with industry partner Ball Aerospace (contract NNM15AA18C).  The Italian contribution is supported by the Italian Space Agency (Agenzia Spaziale Italiana, ASI) through contract ASI-OHBI-2017-12-I.0, agreements ASI-INAF-2017-12-H0 and ASI-INFN-2017.13-H0, and its Space Science Data Center (SSDC) with agreements ASI-INAF-2022-14-HH.0 and ASI-INFN 2021-43-HH.0, and by the Istituto Nazionale di Astrofisica (INAF) and the Istituto Nazionale di Fisica Nucleare (INFN) in Italy.  This research used data products provided by the {\em IXPE} Team (MSFC, SSDC, INAF, and INFN) and distributed with additional software tools by the High-Energy Astrophysics Science Archive Research Center (HEASARC), at NASA Goddard Space Flight Center (GSFC).

The authors thank the referee,  Andrzej Zdziarski, for very detailed and useful comments. J.Pod., M.D., J.S. and V.K. thank for the support from the GACR project 21-06825X and the institutional support from RVO:67985815.
A.V. acknowledges support from the Academy of Finland grants 347003 and 355672.
H.K. and N.R.C. acknowledge NASA support under grants 80NSSC18K0264, 80NSSC22K1291, 80NSSC21K1817, and NNX16AC42G.
P.-O.P. acknowledges financial support from the French High Energy Program (PNHE/CNRS) and from the French space agency (CNES). A.I. acknowledges support from the Royal Society.

\section*{Data Availability}

 The {\em IXPE} data used in this article are publicly available in the HEASARC database (\url{https://heasarc.gsfc.nasa.gov/docs/ixpe/archive/}). The analysis and simulation software \textsc{ixpeobssim} developed by {\em IXPE} collaboration and its documentation is available publicly through the web-page \url{https://ixpeobssim.readthedocs.io/en/latest/?badge=latest.494}. The {NICER} and {\em NuSTAR} data underlying this article are publicly  available from the {\em NuSTAR} (\url{https://heasarc.gsfc.nasa.gov/docs/nustar/archive/nustar_archive.html}) and {NICER} (\url{https://heasarc.gsfc.nasa.gov/docs/nicer/nicer_archive.html}) archives.  The {ART-XC} data used in this article  publicly available at \url{ftp://hea.iki.rssi.ru/public/SRG/ART-XC/data/LMC_X-1/lmc_x_1_barycen.fits.gz}. The \textsc{xspec} software is publicly available in the HEASARC database (\url{https://heasarc.gsfc.nasa.gov/xanadu/xspec/}). The corresponding \textsc{xspec} packages used for this work can be found in the references stated in this article or shared on reasonable request.



\bibliographystyle{mnras}
\bibliography{example} 




\appendix

\section{Observations and data reduction of {NICER}, {\em NuSTAR} and {ART-XC}}
\label{observations_remaining}

\begin{figure*}
    \centering
    \includegraphics[width=\textwidth]{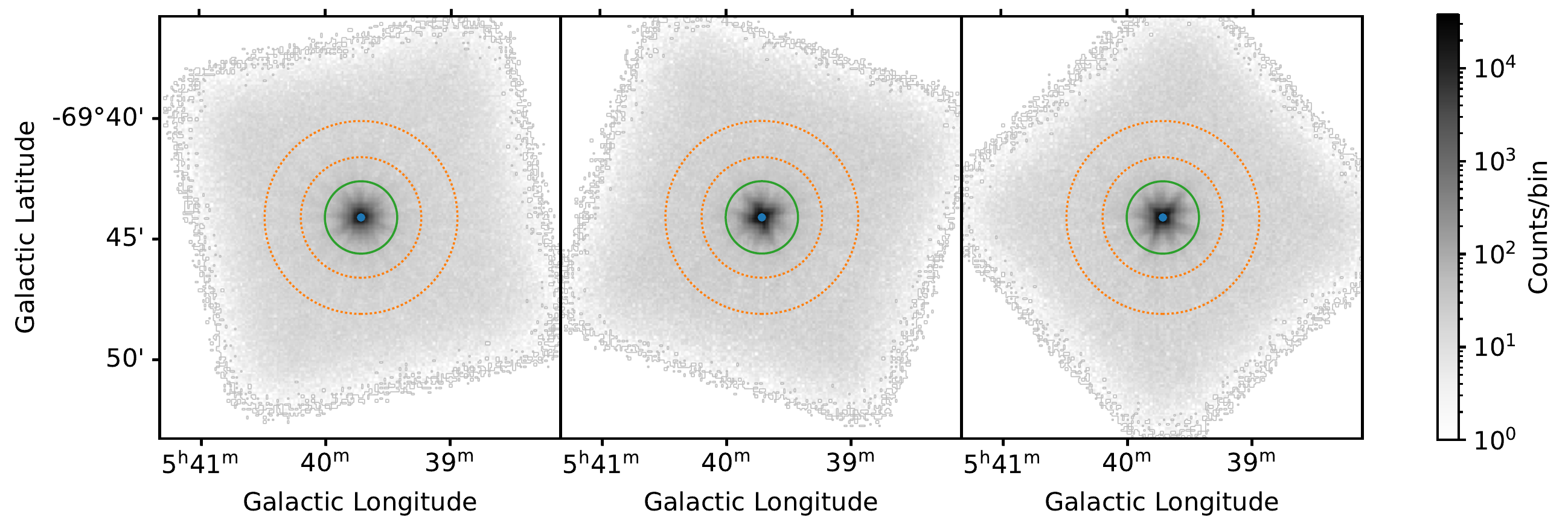}
    \caption{Count maps of the three {\em IXPE} telescopes. The scale of the colour bar is logarithmic to make visible, in addition to the source, also the much fainter background. The regions used to angularly select the source and background in the field of view are the green solid circle and the dashed orange annulus, respectively.}
    \label{fig:counts_map}
\end{figure*}

In Section \ref{observations} we already described in detail the {\em IXPE} data reduction. For completeness, we also show in Fig.~\ref{fig:counts_map} the {\em IXPE} count maps indicating the regions chosen to select the source and the background in the field of view.

We now return to the other three instruments forming the observational campaign. {NICER} \citep{nicer} is a soft X-ray spectral-timing instrument aboard the International Space Station, sensitive within $\sim 0.2$--$12$~keV band. It is non-imaging, and composed of 56 silicon-drift detectors, each of which is paired with a concentrator optic, commonly aligned to a single field approximately 3\arcmin\ in radius. 52 detectors have been active since launch, although in any given observation, some detectors may be temporarily disabled. {NICER} observed LMC X-1 during the course of the {\em IXPE} observational campaign, for a total of 13.5 ks {\em useful} time among 10 ObsIDs from 2022 October 19--28.

{NICER} data were reduced using {\sc nicerl2} with unrestricted undershoot and overshoot rates.  The background was computed using the ``3C50'' model \citep{Remillard_2022}.  Subsequently, the data were filtered to remove intervals with background count rates more than 1 per cent the source rate, and any short GTI intervals <60s were removed.   For each observation, the detectors were screened for outliers in overshoot or undershoot event rates  which are generated by particle background and optical-loading events, respectively.   For both fields, each detector was compared to the detector distribution, and those more than 10$\sigma$ equivalent from the median were filtered out. {NICER} spectra were rebinned in order to oversample the instrumental energy resolution by a factor $\sim 3$.  {NICER} observations were found to be relatively constant in flux and consistent in spectral properties over the {\em IXPE} campaign, and with low power-density rms noise.

The {\em NuSTAR} spacecraft \citep{nustar} acquired a total of $19$ ksec of data on 2022 October 24 under observation ID $90801324002$. The {\em NuSTAR} data were processed with the {\tt NuSTARDAS} software (version 2.1.1) of the {\tt HEAsoft} package (version 6.29) \citep{Heasarc2014}. Source and background events were selected with a circular region of $\sim 67$ arcsec radii for both focal plane modules (FPMA/FPMB). {\tt FTGROUPPHA} was used to rebin the spectra implementing the \cite{Kaastra2016} optimal binning scheme.

We note that we used the cross-calibration model {\tt MBPO} employed in \cite{Krawczynski2022} to reconcile discrepancies between the instruments when performing the joint fits in Sections \ref{spectral_analysis} and \ref{section:Spectro_pol_analysis} with the {NICER}, {\em NuSTAR} and {\em IXPE} spectra. Since the \texttt{KERRBB} normalization is frozen to unity (as should physically be the case) and {\em NuSTAR} has the best absolute flux calibration of the instruments considered here, we fixed the {\em NuSTAR} FPMA normalization to the recommended value of $0.8692$ derived from unfocused observations of the Crab Nebula \citep{Madsen2017}. We allow the normalization of the {\em NuSTAR} FPMB to vary freely. For the NICER observations, we account for cross-calibration discrepancies against {\em NuSTAR} by multiplying the model spectrum with a power law using the same power-law index for all 10 observations and allowing normalization constants to vary. For the {\em IXPE} detector units, all parameters in {\tt MBPO} are allowed to vary freely. See Table \ref{t:mbpoparams} for the best-fitting values of the free parameters in the {\tt MBPO} model for each data set. We also included a $0.5$ per cent systematic uncertainty to all instruments used in the data analysis apart from {NICER}, where we accounted for $1.5$ per cent systematic uncertainty, according to the mission's recommendation.\footnote{Available at \url{https://heasarc.gsfc.nasa.gov/docs/nicer/analysis_threads/cal-recommend/}.} This is necessary to take into account the unknown internal calibration.
\renewcommand{\thetable}{3}
\begin{table}
\begin{center}
\resizebox{\columnwidth}{!}{
\begin{tabular}{p{2cm}p{2cm}p{2cm}p{2cm}}
\hline
\hline
Data & {\tt MBPO} Parameter (unit) & Description & Value\\  
\hline
\hline
{NICER} 1 & $\Delta \Gamma_1$ & Power-law index & $-0.153 \pm 0.008$ \\
& norm & Normalization & $0.879 \pm 0.005$ \\
{NICER} 2 & norm & Normalization & $0.896 \pm 0.005$ \\ 
{NICER} 3 & norm & Normalization & $0.840 \pm 0.004$ \\ 
{NICER} 4 & norm & Normalization & $1.022 \pm 0.007$ \\ 
{NICER} 5 & norm & Normalization & $0.958 \pm 0.005$ \\ 
{NICER} 6 & norm & Normalization & $0.967 \pm 0.005$ \\ 
{NICER} 7 & norm & Normalization & $1.000 \pm 0.005$ \\ 
{NICER} 8 & norm & Normalization & $0.944 \pm 0.004$ \\ 
{NICER} 9 & norm & Normalization & $0.864 \pm 0.005$ \\ 
{NICER} 10 & norm & Normalization & $0.879 \pm 0.005$ \\ 
\hline
{\em NuSTAR} FPMB & norm & Normalization & $0.867 \pm 0.005$ \\ 
\hline
{\em IXPE} DU1 & $\Delta \Gamma_1$ & Low-energy power-law index & $-0.296 \pm 0.009$ \\ 
& $\Delta \Gamma_2$ & High-energy power-law index & $1.1 \pm 0.3$ \\ 
& $E_\textrm{br}$ (keV) & Break energy & $6.38 \pm 0.03$ \\ 
& norm & Normalization & $0.693 \pm 0.005$ \\ 
{\em IXPE} DU2 & $\Delta \Gamma_1$ & Low-energy power-law index & $-0.254 \pm 0.009$ \\ 
& $\Delta \Gamma_2$ & High-energy power-law index & $1.3 \pm 0.7$ \\ 
& $E_\textrm{br}$ (keV) & Break energy & $6.77 \pm 0.04$ \\ 
& norm & Normalization & $0.684 \pm 0.005$ \\ 
{\em IXPE} DU3 & $\Delta \Gamma_1$ & Low-energy power-law index & $-0.247 \pm 0.009$ \\ 
& $\Delta \Gamma_2$ & High-energy power-law index & $1.6 \pm 0.4$ \\ 
& $E_\textrm{br}$ (keV) & Break energy & $6.49 \pm 0.04$ \\ 
& norm & Normalization & $0.660 \pm 0.005$ \\
\hline
\end{tabular}
}
\caption{Best-fitting free parameters (with uncertainties at 90 per cent confidence level) of the cross-calibration model {\tt MBPO} employed on the {NICER}, {\em NuSTAR} and {\em IXPE} data for the fit presented in Table \ref{t:spectralfitparams}. See text for details.}
\label{t:mbpoparams}
\end{center}
\end{table}

The {Mikhail Pavlinsky} {ART-XC} telescope is a grazing incidence focusing X-ray telescope \citep{2021A&A...650A..42P} on board the \textit{Spectrum-Rontgen-Gamma observatory} ({\em SRG}, \citealt{2021A&A...656A.132S}). It observed LMC X-1 on 2022 Oct 27 with a total exposure of 84.4 ks. 
The {ART-XC} observation has two short technical interruptions of $\sim 100$ s duration each. 
{ART-XC} data were processed with the analysis software {\tt ARTPRODUCTSv1.0} and the {\tt CALDB}  (calibration data base) version 20220908.

\section{Spectral fit with {\tt COMPPS} model}\label{compps_fit}

To use consistent spectral parameters in the polarimetric modeling of Section \ref{compps_modelling}, we fitted the spectra from {NICER} and {\em NuSTAR} with the same model, {\tt COMPPS}, in slab geometry \citep{PS1996} with thermal distribution of the electrons, and with multi-temperature blackbody emission from the disc as seed photons. The spectra used are the same as in Section \ref{spectral_analysis}. We used an energy range of 1--8 keV for {NICER} and 3--20 keV for {\em NuSTAR}. The spectra above 20 keV were background dominated in {\em NuSTAR}. To prevent potential confounding effects arising from the soft excess feature discussed in Section \ref{spectral_analysis}, data below 1 keV for {NICER} were excluded from the analysis. In the {\tt COMPPS} model, the covering fraction was fixed to unity while the reflection fraction was set to zero, as no reflection features were seen in the spectra. We used a constant to account for instrumental uncertainties, {\tt TBFEO} for the neutral absorption and an additional gaussian absorption between 9 to 10 keV {\tt GABS} to account for an absorption feature discussed in detail in the spectral analysis in Section \ref{spectral_analysis}. The constant for {NICER} spectrum was frozen at $1.0$, while the fit resulted in $0.91$ and $0.89$ for {\em NuSTAR}-FPMA and {\em NuSTAR}-FPMB, respectively. We obtained the best fit for an inner disc temperature of $0.81\pm0.01$ keV, for an electron temperature of $10\pm1$ keV and for an optical depth of $1.26\pm0.09$. The $\chi^2/{\rm dof}$ for the fit is $1231/1125$. The parameters of the fit are outlined in Table \ref{t:spectralfitcompps}. The $\chi^2/{\rm dof}$ appears better than for the spectral fit described in Section \ref{spectral_analysis} due to the intentional reduction of the {NICER} energy range, which excludes some intricate spectral features in the soft X-rays (see Fig. \ref{fig:spectralfit}), and due to better capture of the {\tt COMPPS} model of the joint spectrum. However, we do not keep this spectral fitting attempt as leading in the main paper body, because the {\tt COMPPS} model cannot separate the thermal and Comptonized component, which is necessary for the basic polarization analysis of the two dominant components performed in Section \ref{section:Spectro_pol_analysis}.

\renewcommand{\thetable}{4}
\begin{table}
\begin{center}
\resizebox{\columnwidth}{!}{
\begin{tabular}{p{1.5cm}p{2cm}p{2.5cm}p{2cm}}
\hline
\hline
Component & Parameter (unit) & Description & Value\\  
\hline
\hline
{\tt TBFEO} & $N_\textrm{H}$ ($10^{22}$\,cm$^{-2}$) & Hydrogen column density &  $0.624_{-0.165}^{+0.003}$ \\
& O & Oxygen abundance &  $0.82^{+0.01}_{-0.44}$ \\
& Fe & Iron abundance & $<0.41$\\
& $z$ & Redshift & $0.0$ (frozen)\\
\hline
{\tt COMPPS} & $\tau$ & Optical depth & $1.26\pm0.09$ \\
& $kT_\textrm{e}$ (keV) & Electron temperature & $10\pm1$ \\
& $kT_\textrm{bb}$ (keV) & Inner disc temperature & $0.81\pm0.01$\\
& cosIncl & Cosine of the inclination angle & $0.81$ (frozen)  \\
& cov\_fac & Covering fraction & 1 (frozen) \\
& $R$ & Reflection fraction & 0 (frozen) \\
& norm & Normalization & $94\pm4$\\ 
\hline
{\tt GABS} & $E_\textrm{l}$ (keV) & Line energy &  $9.64$ (frozen) \\
& $\sigma$ (keV) & Line width & $1.17$ (frozen)\\
& Strength (keV) & Line depth & $1.15$ (frozen) \\
\hline
\end{tabular}
}
\caption{Parameters of the best fit (with uncertainties at 90 per cent confidence level) of the joint {NICER} and {\em NuSTAR} spectra using the {\tt COMPPS} model. $G_\textrm{min}$ parameter of the {\tt COMPPS} was set to $-1$ to obtain a fully thermal distribution of electrons, and all other parameters not mentioned in the table are set at default values.
$\chi^2/{\rm dof}$ for the fit is $1231/1125$.}
\label{t:spectralfitcompps}
\end{center}
\end{table}



\section*{Authors' affiliations}
\footnotesize{
     $^{1}$ Université de Strasbourg, CNRS, Observatoire Astronomique de Strasbourg, UMR 7550, 67000 Strasbourg, France \\
     $^{2}$ Astronomical Institute of the Czech Academy of Sciences, Bo{\v c}ní II 1401/1, 14100 Praha 4, Czech Republic \\
     $^{3}$ Astronomical Institute, Charles University, V Hole{\v s}ovi{\v c}kách 2, CZ-18000 Prague, Czech Republic \\
     $^{4}$ Dipartimento di Matematica e Fisica, Università degli Studi Roma Tre, Via della Vasca Navale 84, 00146 Roma, Italy \\
     $^{5}$ INAF Istituto di Astrofisica e Planetologia Spaziali, Via del Fosso del Cavaliere 100, I-00133 Roma, Italy\\
    $^{6}$ Physics Department and McDonnell Center for the Space Sciences, Washington University in St. Louis, St. Louis, MO 63130, USA \\
    $^{7}$ Center for Astrophysics | Harvard \& Smithsonian, 60 Garden St, Cambridge, MA 02138, USA \\
    $^{8}$ Department of Physics and Astronomy, FI-20014 University of Turku, Finland \\
     $^{9}$ Nordita, KTH Royal Institute of Technology and Stockholm University, Hannes Alfvéns väg 12, SE-10691 Stockholm, Sweden \\     
     $^{10}$ NASA Marshall Space Flight Center, Huntsville, AL 35812, USA \\
    $^{11}$ California Institute of Technology, Pasadena, CA 91125, USA\\
    $^{12}$ Université Grenoble Alpes, CNRS, IPAG, 38000 Grenoble, France \\
    $^{13}$ Space Research Institute of the Russian Academy of Sciences, Profsoyuznaya Str. 84/32, Moscow 117997, Russia \\
     $^{14}$ Department of Astronomy, University of Maryland, College Park, Maryland 20742, USA \\
    $^{15}$ Center for Research and Exploration in Space Science and Technology, NASA/GSFC, Greenbelt, MD 20771, USA \\
     $^{16}$ School of Mathematics, Statistics, and Physics, Newcastle University, Newcastle upon Tyne NE1 7RU, UK \\
    $^{17}$ Hiroshima Astrophysical Science Center, Hiroshima University, 1- 3-1 Kagamiyama, Higashi-Hiroshima, Hiroshima 739-8526, Japan \\
     $^{18}$ Dipartimento di Fisica, Università degli Studi di Roma “Tor Vergata”, Via della Ricerca    Scientifica 1, I-00133 Roma, Italy\\
     $^{19}$ Istituto Nazionale di Fisica Nucleare, Sezione di Roma "Tor Vergata", Via della Ricerca Scientifica 1, 00133 Roma, Italy \\
    $^{20}$ Mullard Space Science Laboratory, University College London, Holmbury St Mary, Dorking, Surrey RH5 6NT, UK \\
    $^{21}$ Instituto de Astrofísica de Andalucía—CSIC, Glorieta de la Astronomía s/n, 18008 Granada, Spain \\
    $^{22}$ INAF Osservatorio Astronomico di Roma, Via Frascati 33, 00078 Monte Porzio Catone (RM), Italy \\
     $^{23}$ Space Science Data Center, Agenzia Spaziale Italiana, Via del Politecnico snc, 00133 Roma, Italy \\
    $^{24}$ INAF Osservatorio Astronomico di Cagliari, Via della Scienza 5, 09047 Selargius (CA), Italy \\
    $^{25}$ Istituto Nazionale di Fisica Nucleare, Sezione di Pisa, Largo B. Pontecorvo 3, 56127 Pisa, Italy \\
    $^{26}$ Dipartimento di Fisica, Università di Pisa, Largo B. Pontecorvo 3, 56127 Pisa, Italy \\
    $^{27}$ Istituto Nazionale di Fisica Nucleare, Sezione di Torino, Via Pietro Giuria 1, 10125 Torino, Italy \\
    $^{28}$ Dipartimento di Fisica, Università degli Studi di Torino, Via Pietro Giuria 1, 10125 Torino, Italy \\
    $^{29}$ INAF Osservatorio Astrofisico di Arcetri, Largo Enrico Fermi 5, 50125 Firenze, Italy \\
    $^{30}$ Dipartimento di Fisica e Astronomia, Università degli Studi di Firenze, Via Sansone 1, 50019 Sesto Fiorentino (FI), Italy \\
    $^{31}$ Istituto Nazionale di Fisica Nucleare, Sezione di Firenze, Via Sansone 1, 50019 Sesto Fiorentino (FI), Italy \\
     $^{32}$ ASI - Agenzia Spaziale Italiana, Via del Politecnico snc, 00133 Roma, Italy \\
     $^{33}$ Science and Technology Institute, Universities Space Research Association, Huntsville, AL 35805, USA\\
    $^{34}$ Department of Physics and Kavli Institute for Particle Astrophysics and Cosmology, Stanford University, Stanford, California 94305, USA \\
    $^{35}$ Institut für Astronomie und Astrophysik, Universität Tübingen, Sand 1, 72076 Tübingen, Germany \\
    $^{36}$ RIKEN Cluster for Pioneering Research, 2-1 Hirosawa, Wako, Saitama 351-0198, Japan\\
    $^{37}$ Yamagata University, 1-4-12 Kojirakawamachi, Yamagatashi 990-8560, Japan \\
    $^{38}$ Osaka University, 1-1 Yamadaoka, Suita, Osaka 565-0871, Japan\\
    $^{39}$ University of British Columbia, Vancouver, BC V6T 1Z4, Canada \\
    $^{40}$ International Center for Hadron Astrophysics, Chiba University, Chiba 263-8522, Japan\\
    $^{41}$ Institute for Astrophysical Research, Boston University, 725 Commonwealth Avenue, Boston, MA 02215, USA \\
    $^{42}$ Department of Astrophysics, St. Petersburg State University, Uni- versitetsky pr. 28, Petrodvoretz, 198504 St. Petersburg, Russia \\
    $^{43}$ Department of Physics and Astronomy and Space Science Center, University of New Hampshire, Durham, NH 03824, USA\\
    $^{44}$ Finnish Centre for Astronomy with ESO, 20014 University of Turku, Finland \\
    $^{45}$ Istituto Nazionale di Fisica Nucleare, Sezione di Napoli, Strada Comunale Cinthia, 80126 Napoli, Italy\\
    $^{46}$ MIT Kavli Institute for Astrophysics and Space Research, Massachusetts Institute of Technology, 77 Massachusetts Avenue, Cambridge, MA 02139, USA \\
    $^{47}$ Graduate School of Science, Division of Particle and Astrophysical Science, Nagoya University, Furocho, Chikusaku, Nagoya, Aichi 464-8602, Japan \\
    $^{48}$ Department of Physics, The University of Hong Kong, Pokfulam, Hong Kong \\ 
    $^{49}$ Department of Astronomy and Astrophysics, Pennsylvania State University, University Park, PA 16802, USA \\
    $^{50}$ INAF Osservatorio Astronomico di Brera, Via E. Bianchi 46, 23807 Merate (LC), Italy \\
    $^{51}$ Dipartimento di Fisica e Astronomia, Università degli Studi di Padova, Via Marzolo 8, 35131 Padova, Italy \\
    $^{52}$ Anton Pannekoek Institute for Astronomy \& GRAPPA, University of Amsterdam, Science Park 904, 1098 XH Amsterdam, The Netherlands \\
    $^{53}$ Guangxi Key Laboratory for Relativistic Astrophysics, School of Physical Science and Technology, Guangxi University, Nanning 530004, China \\
}

\bsp	
\label{lastpage}
\end{document}